\documentclass[12pt]{iopart}

\usepackage{color}

\usepackage{graphicx}

\newcommand{\TMYAG}{Tm$^{3+}$:YAG }

\begin{document}
\title{Atomic Frequency Comb storage as a slow-light effect}
\author{M. Bonarota$^1$, J.-L. Le~Gou\"et $^1$, S. A. Moiseev $^{2,3}$, T. Chaneli\`ere$^1$}
\address{$^1$ Laboratoire Aim\'e Cotton, CNRS-UPR 3321, Univ. Paris-Sud, B\^at. 505, 91405 Orsay cedex, France}

\address{$^2$ Kazan Physical-Technical Institute of the Russian Academy of Sciences,10/7 Sibirsky Trakt, Kazan, 420029, Russia}
\address{$^3$ Institute for Informatics of Tatarstan Academy of Sciences, 20 Mushtary, Kazan, 420012, Russia}

\ead{thierry.chaneliere@u-psud.fr}

\begin{abstract}
Atomic Frequency Comb (AFC) protocol has been particularly successful recently to demonstrate the storage of quantum information in a solid medium (rare-earth doped crystals).  The AFC is inspired by the photon-echo technique. We show in this paper that the AFC is actually closely related to the slow-light based storage protocols extensively used in atomic vapours. Experimental verifications  are performed in thulium doped YAG (\TMYAG). We clarify the interplay between absorption and dispersion and propose a classification of the existing protocols.

\end{abstract}
\submitto{\jpb}
\maketitle

\section{Introduction}
The prospect of storing quantum states of light is based on very different approaches in terms of systems and protocols. Slow-light produced by electromagnetically induced transparency (EIT) in atomic vapours is now a well-known phenomenon and can be considered as an archetype or at least as an historical precursor. EIT opens a transparency window inducing a strong dispersion profile where the group velocity can be much lower than the  speed of light (slow-light). It is sufficient to map the incoming signal into the atomic excitation (spin wave). The strong Raman field used to control the coupling between the signal and the spin wave is simply switched off to keep the excitation freezed in the medium.

More recently protocols inspired by the photon-echo technique \cite{Elyutin} have attracted a lot of attention relying on the solid conceptual basis of the NMR context \cite{anderson:1324}. The interest then was also triggered by the consideration of solid state materials namely rare-earth doped crystals (REDC) known for their long coherence time at low temperature. Their classical light storage capacity has been studied for decades \cite{Mossberg:82,mitsunaga1990248, lin1995demonstration}, so it was quite natural to reconsider REDC with a quantum point of view.  Unfortunately the strong $\pi$ rephasing pulse would practically disturb the detectors and fundamentally induce noise by spontaneous emission \cite{Sangouard2PE, Ruggiero}. Aware of this dead-end, one rapidly proposed alternative protocols  still inspired by the PE (i.e. based on coherent transients) but exempt from their limitations.

The controlled reversible inhomogeneous broadening (CRIB) is a good example. No strong rephasing pulse is used. The echo emission can be interpreted as a reversal of the absorption process. The reversibility is achieved by a direct control of the atomic detuning \cite{Moiseev2001, LAPL:LAPL200310071}.  The Stark effect offers this possibility. A static electrical field as an external command does the rephasing by controlling the dipole frequencies within the inhomogeneous broadening \cite[and references therein]{tittel-photon}. In the same context the AFC has been proposed \cite{AFCTh}. It works in Stark insensitive material and offers a larger temporal multiplexing capacity \cite{nunn}. It also allowed very recently the demonstration of entanglement storage in a crystal \cite{Clausen2011,Saglamyurek2011}. As in CRIB, no strong pulse is used. The absorption profile is initially prepared by spectrally selective optical pumping (frequency comb like structure). The signal is then diffracted off a spectral grating producing an echo in the time-domain. The echo can be stored for a longer time by a Raman transfer toward hyperfine states \cite{Moiseev2001,AFCspin}.

From the present analysis, one could conclude that within a wide landscape of systems and protocols, we have on one side the EIT in atomic vapours and on the other side the PE relatives in REDC. This vision is incorrect. The goal of this paper is to show that even if the AFC is clearly inspired by the PE, it is actually based on slow-light and is then similar to EIT.

To raise the doubts, we can briefly compare AFC and CRIB. AFC and CRIB are fundamentally as different as the three-pulse and two-pulse PE are. For the two-pulse PE (2PE) and CRIB, the signal is totally absorbed in the medium and stored as coherences. The retrieval is triggered by a $\pi$ pulse and a static electric  field respectively. For the three-pulse PE (3PE) and AFC, the signal is directly diffracted by a spectral population grating to produce an echo in the time domain. For both, the population grating is prepared before the arrival of the signal. No information is prerecorded at the preparation stage except the retrieval time of the future incoming signal whose temporal shape can be arbitrary. The AFC can actually be seen as an extreme situation of accumulated 3PE. The periodic structure is composed of narrow absorbing peaks (atomic comb).

As a short-cut, we could say that AFC is so different than CRIB that it actually resembles EIT and related slow-light protocols \cite{lauroslow}. To support this statement we consider slow-light effects in AFC. We first show that dispersion effects should be dominant to obtain a large storage efficiency. An interpretation in terms of group velocity and delay is then possible. To investigate this regime, we perform a series of experiments in \TMYAG. To conclude our analysis, we propose a classification of the different protocols regardless of the system they are applied in.

\section{Dispersion Vs Absorption in AFC}
\begin{figure}[ht]
	\begin{center}
	\includegraphics[width=12cm]{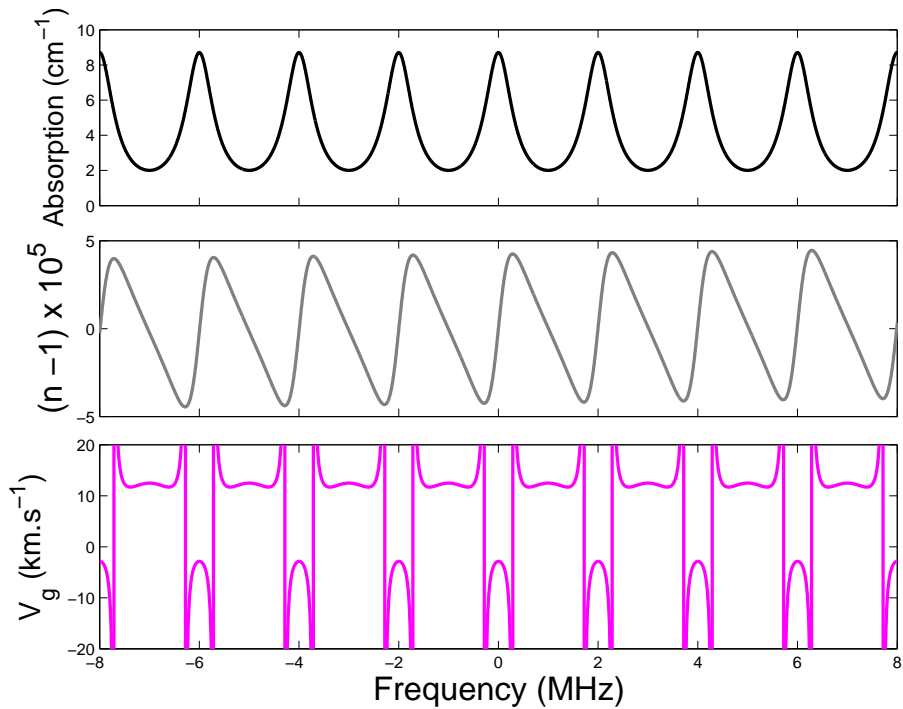}
	
	\caption{  {Atomic frequency comb composed of lorentzian peaks whose half-width at half-maximum  (HWHM) is a third of the comb spacing (see section \ref{effAFC} for a detailed definition). The maximum absorption (top) correspond to a 5mm \TMYAG crystal. The group velocity $V_g$ (bottom) is a more significant metric of the dispersion than the bare refractive index n  (middle). We simply use the general formula $\displaystyle V_g =\frac{c}{n+ \omega\left( dn/d\omega\right)}$.}}
\label{figDisp}	\end{center}
\end{figure}

 {Storing light represents an efficient interaction by the excitation of the atomic coherences. Considering this superposition state is also a possible grid of analysis. On resonance excitation leading to absorption is a straightforward manner to produce a superposition state. Off-resonance field can also produce a transient atomic excitation of the coherences known as adiabatic following \cite{Grischkowsky, lauroslow, allen1987ora}. It is an efficient energy transfer between field and atoms as well.}  {Here we would like to separate dispersion and absorption. It is always a bit artificial since both are connected by the Kramers-Kronig relation. To start our analysis, we propose to clarify this notions. An absorbing periodic structure composed of lorentzian peaks allows analytical calculation of the efficiency \cite{AFC_NJP} and will be considered here. It is then relatively easy to calculate the refraction index (dispersion) by starting with a complex lorentzian {\it de facto} satisfying the Kramers-Kronig relation. The real part gives the absorption and the imaginary part the refraction index respectively or {\it vice versa} if the suseptibility is considered (see fig. \ref{figDisp}).  We take realistic values for the maximum absorption and the comb spacing typically corresponding to our experimental measurement as we will see later on.}

 {Fig. \ref{figDisp} is the basis of our qualitative analysis. The group velocity shows two distinct behaviours. Between the absorbing peaks, the group velocity is reduced (slow-light) and positive (normal dispersion). On resonance with the peaks, dispersion is anomalous (negative group velocity). Anomalous dispersion is difficult to observed because the dominant effect is a reduction of the intensity in this region \cite{Garrett,Akulshin,Dogariu}. In other words, the propagation is dominated by the absorption. These two regions of the spectrum represent two different propagations effects: normal dispersion (simply called dispersion by now)  and absorption.}

More precisely, dispersion dominates when the incoming spectrum is essentially far-off-resonance. The pulses is distorted without loss of energy to the first order. Slow-light is just a particular case of linear dispersion producing a constant group delay.

Absorption takes place close to resonance. The pulse losses part of its energy to directly promote the atoms to the excited state. It is responsible for the well-known Bouguer-Lambert-Beer absorption law \cite{crisp1970psa}.

In the AFC situation, the incoming spectrum covers many narrow absorbing peaks. One could conclude that dispersion and absorption coexist: dispersion for the spectral components in between the  peaks and absorption close to the peaks. We here show that for large AFC retrieval efficiencies, dispersion should dominate. This point has been partially addressed in \cite{AFC_PRA} where a systematic study of the efficiency optimization has been done. We reconsider the situation in the context of slow-light. If dispersion is dominant, most of the incoming energy should be transmitted. This statement is made obvious when the propagation of the energy spectral density is considered. We here write it as $ |\widetilde {\Omega}\left(z,\omega \right)|^2$ where ${\Omega}\left(z,t \right)$ is the time-dependent Rabi frequency. The formalism of the Bloch-Maxwell equations in the slowly varying amplitude and the rotating wave approximations \cite{allen1987ora} will be used later on. In the weak signal limit \cite{crisp1970psa}, the propagation of the energy spectral density follows the Bouguer-Lambert-Beer law:

\begin{equation}\label{energy_propag}
\partial_z |\widetilde {\Omega}\left(z,\omega \right)|^2= - \alpha\left(\omega\right)  |\widetilde {\Omega}\left(z,\omega \right)|^2
\end{equation}

The inhomogeneous broadening is written in the form of a frequency dependent absorption coefficient $\alpha\left(\omega\right)$. We here assume that the homogeneous broadening $\gamma$ is negligible (otherwise it would appear as a convolution with $\alpha\left(\omega\right)$ \cite{AFC_PRA}). 

\subsection{How to make an efficient AFC ?}\label{effAFC}

 {We again consider an absorbing periodic structure composed of well-separated lorentzian peaks \cite{AFC_NJP}.} The way to optimize the efficiency is nevertheless very general. The highest optical depth is desirable. As the optical depth is increasing, the width of the peaks, whatever is their exact shape should be reduced to optimize the efficiency.

More precisely for lorentzian peaks ( $\Gamma$ is the HWHM) with an optical depth $\alpha_M L$ corresponding to the absorption maxima ($L$ is the length of the medium), the efficiency is

\begin{equation}\label{effLorentz}
\eta=\left(\alpha_M L\right)^2 \left(\Gamma T\right)^2 e^{-\Gamma T \left(2+\alpha_M L/2\right)}
\end{equation}
in the limit of a large optical depth \cite{AFC_NJP}.  $2\pi/T$ is the comb spacing in the spectral domain. We see that for a given absorption $\alpha_M L$, the width of the peak should be adjusted to the optimal width $\displaystyle \Gamma_{opt}\approx \frac{1}{T} \frac{4}{\alpha_M L}$ maximizing the efficiency. In this limit, the efficiency increases asymptotically toward the maximum, 54\% in the forward direction following the formula  \cite{AFC_NJP}

\begin{equation}\label{effopt}
\eta_{opt}=4e^{-2} \left(\alpha_M L\right)^2/\left({4+\alpha_M L}\right)^2
\end{equation}

\subsection{For an optimal AFC, what is the transmitted energy ?}
When the width of the peaks is effectively adjusted to $ \Gamma_{opt}$ optimizing the efficiency for an increasing absorption, one can consider the transmission of the energy spectral density by integrating eq. \ref{energy_propag}.

In the vicinity of an absorption peak, the transmission coefficient is
\begin{equation}\label{T_lorentz}
T\left(\omega \right) \simeq \exp \left(-\frac{\alpha_M L}{1+4\omega^2/\Gamma_{opt}^2}\right)
\end{equation}
The energy spectral density is significantly absorbed within a range given by $\Gamma_{opt} \sqrt{\alpha_M L}$. To scale the absorbed energy, one can simply compare the absorption window and the comb spacing. The absorbed fraction is then roughly given by  $\Gamma_{opt} T \sqrt{\alpha_M L}$ which scales as $\displaystyle \frac{1}{ \sqrt{\alpha_M L}}$. It goes to zero when the absorption is increased.

To summarize, it tells us that when the peak width is adjusted to maximize the efficiency, for an increasing optical depth the energy is completely transmitted. It doesn't mean that the incoming pulse is not modified. The pulse is distorted but the total energy is retrieved after integration. In other words, the production of an echo (multiple echos eventually) is a special kind of distortion but it does not leave energy behind. In that sense, the AFC retrieval is a pure dispersive effect.

The particular example of lorentzian peaks can be generalized because the exact peak shape is not really critical in the high efficiency regime  \cite{AFC_PRA}. This is precisely because the production of the echo is dominated by off-resonant excitation of the comb peaks. This argument confirms the importance of dispersion as opposed to absorption.

The ideal situation of very narrow and absorbing peaks is hard to achieved in practice. It requires an extremely large optical depth to start with and an accurate tailoring of the population to prepare the comb. In order to investigate the dispersive regime, we take an alternative approach. We consider the diffraction of an incoming pulse train. 

\section{AFC echo from a single pulse to a pulse train}

A single pulse uniformly covers many comb peaks. On the contrary, an incoming train whose pulse separation matches the inverse comb spacing, has an optical spectrum restricted to the transparency regions of the comb. In that case, the effect should be essentially dispersive. We first investigate the AFC echo of a single pulse. It corresponds to the situation of pulse storage \cite{AFCTh}. We then consider the diffraction of an increasing number of pulses making our interpretation in terms of slow-light more clear. The group delay is finally measured.

\subsection{Echo of a single pulse}

This situation has been extensively described previously. We here briefly review it. We also use a novel preparation method that deserves further explanation.

\subsection{Preparation method}

Experiments are conducted in a 0.5\%-doped \TMYAG interacting at 793 nm. The sample is maintained at low temperature (2.3K) in a liquid helium cryostat ensuring a narrow homogeneous linewidth (10 kHz) making possible the tailoring of a fine absorbing spectral structure. A 40G magnetic field is applied along the [001] crystalline axis parallel to the polarization axis. It splits the ground and excited states into two Zeeman sublevels ($\Delta_e = 240kHz$ and $\Delta_g = 392kHz$)\cite{split}. The frequency comb is obtained by spectral hole burning (frequency-selective optical pumping), using Zeeman ground sublevels as shelving states. Its long lifetime (more than 1s at 2.3K \cite{Ohlsson}) allows a long-lived and well-contrasted comb.

 {The presence of the Zeeman structure may be a source of confusion. Nevertheless in the weak signal limit the different transitions can be considered as independent. The situation would be drastically different if a $\pi$ Raman transfer was applied as in the complete AFC protocol \cite{AFCTh}. We here restrict ourselves to the primary stage where the signal in diffracted in the forward direction by a population grating. So we simply model the structure by a frequency dependent absorption. }

The square-shaped comb is known to optimize the echo efficiency for a given initial optical depth.  A specific preparation sequence involving weak monochromatic pulses was used to obtain a square like structure \cite{AFC_PRA}. We here propose to relying on Complex Hyperbolic Secants (CHS) solely. A CHS is a chirped light pulse whose complex amplitude is defined by : $$ \Omega(t) = \Omega_0 \mathrm{sech}\left(\beta t\right)^{1-i\mu} $$ where $\beta $ is inversely proportional to the pulse duration and $ \mu $ is a positive constant \cite{De_Seze, Hoult}.  {The frequency is swept between $\omega_0 - \mu \beta$ and  $\omega_0 + \mu \beta$ where $\omega_0$ is the carrier frequency}.

Directly inspired by NMR composite pulses \cite{Hoult}, CHS are more robust against experimental artefacts (power fluctuations for example). Their are also powerful alternatives to $\pi$-pulses in the PE sequence \cite{ROSE}. Under appropriate conditions such a pulse inverts the atomic population over a sharply defined rectangular frequency range of extension $ 2\beta\mu $ around the central frequency within the inhomogeneous profile \cite{De_Seze, roos}. Spano {\it et al.} \cite{Warren} showed that the inversion can also be conserved in optically thick media rendering the CHS pulses robust against propagation effects. This is precisely what is desired for the preparation sequence. So, to burn a square-shaped composed of 2n+1 transmission peaks separated by a frequency step $2 \pi /T$ we apply a pattern of 2n+1 CHSs each centered on $\omega_0 - 2 \pi k /T$ ($k=-n,-n+1, ...,0, ...,n$) followed by a 5 ms dead-time ensuring the decay from the excited state (lifetime $800 \mu s$).  {CHS pulses are separated by 1 $\mu$s that should not be confused by T. Here the periodic spacing in the spectral domain is solely given by the central frequencies of the CHS pulses.} This pumping scheme is repeated to achieve the population stationary regime where a well-contrasted square-shaped comb is burnt in the atomic population. Pulses are controlled in amplitude and frequency by acousto-optic modulators driven by an arbitrary waveform generator.

\begin{figure}[ht]
	\begin{center}
	\includegraphics[width=15cm]{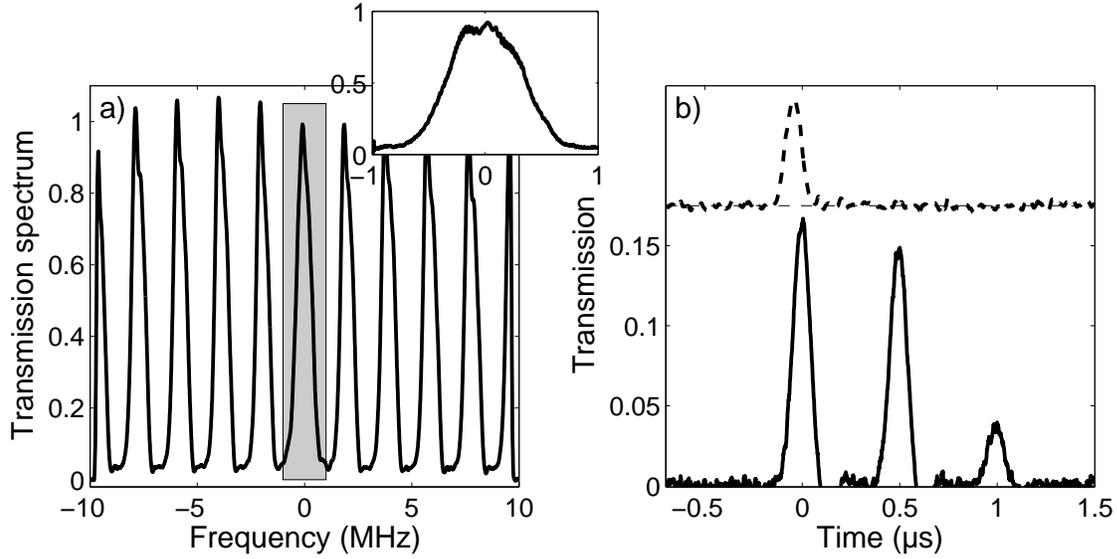}
	
	\caption{a).Complete AFC prepared with a series of CHSs. The period is $1 /T = 2MHz$.  {The probe laser is swept across 20 MHz in 200 $\mu s$ to observe AFC peaks. It is slightly too rapid and produces ring-downs on the narrow peaks. It explains the fact that the transmission looks larger than one. It is actually not the case as we see in the inset showing the central peak recorded with a slower sweeping rate to prevent any distortion.} To obtain a well-contrasted comb, we choose $\beta = 120kHz$ and $\mu = 2$. The resulting peaks are typically 400kHz wide. b). The dashed line represents the incoming pulse. It is partially transmitted and gives rise to a retrieval with an efficiency of about 15\% (solid line).}
\label{peig}	\end{center}
\end{figure}

A very weak probe beam is swept to record the absorption spectrum (fig.\ref{peig}.a) (the probe beam is twice smaller than the pumping one).

We use counterpropagating beams in order to prevent our probe detector (avalanche photo-diode) to be exposed to the strong pumping field. An even better rejection is obtained by using crossed polarized beams and a polarization beamsplitter before the detector. This polarization configuration is made possible by the cubic arrangement of the active ions in the matrix \cite{AFC_FM}.

\subsection{AFC echo in \TMYAG} \label{sectionEcho}

To validate our preparation method, we then send a single gaussian probe pulse. It is diffracted by the comb. Its duration is fixed to 100ns covering several peaks of the AFC. The first echo is retrieved after $T=500$ns (fig.\ref{peig}.b). 

We observe an efficiency of nearly 15\% consistent with our previous observations in the same material \cite{AFC_PRA, AFC_NJP}.

The efficiency can be calculated from the absorption spectrum as previously explained in different publications \cite{ AFC_NJP, AFCsingle, Sonajalg:94}.

Using the Bloch-Maxwell formalism, one can derive the efficiency by assuming a fully periodic absorption spectrum  $\displaystyle \alpha\left(\Delta\right)$. Its decomposition in a Fourier series $\displaystyle \alpha\left(\Delta\right)= \sum_n \alpha_n e^{i n \Delta T}$ allows the calculation of the successive echos amplitude.

The first echo intensity defining the protocol efficiency is 

\begin{equation}\label{etag0g1}
\eta\left( L\right)  =\displaystyle \left|\alpha_{1} L \right|^2  e^{-\alpha_{0} L }
\end{equation}

It should be noted that this formula is more general than eq. \ref{effLorentz}. The later is a particular case of eq. \ref{etag0g1} assuming a lorentzian shape of the peaks developed in a Fourier series. 

From the experimental absorption spectrum $\displaystyle \alpha \left(\Delta\right)$ (fig.\ref{peig}.a), we can numerically  calculate $\alpha_{0}$ and $\alpha_{1}$ by using the general formula $\displaystyle \alpha_n = \frac{T}{2 \pi} \int_{-\pi/T}^{\pi/T} \mathrm{d} \Delta \, \alpha\left(\Delta\right) e^{-i n \Delta T}$. There is no specific assumption one the exact peaks shape. Following this procedure, we expect an efficiency of  14\%  to be compared with the 15\% actually measured for a single incoming pulse. The slight discrepancy cannot be considered as significant because it is relatively difficult to accurately measure high optical depths \cite{AFC_NJP}.

By increasing the number of incoming pulses, one can analyze the role of dispersion in the echo generation.

\subsection{Echo of few pulses: intermediate regime}

\begin{figure}[ht]
	\begin{center}
	\includegraphics[width=15cm]{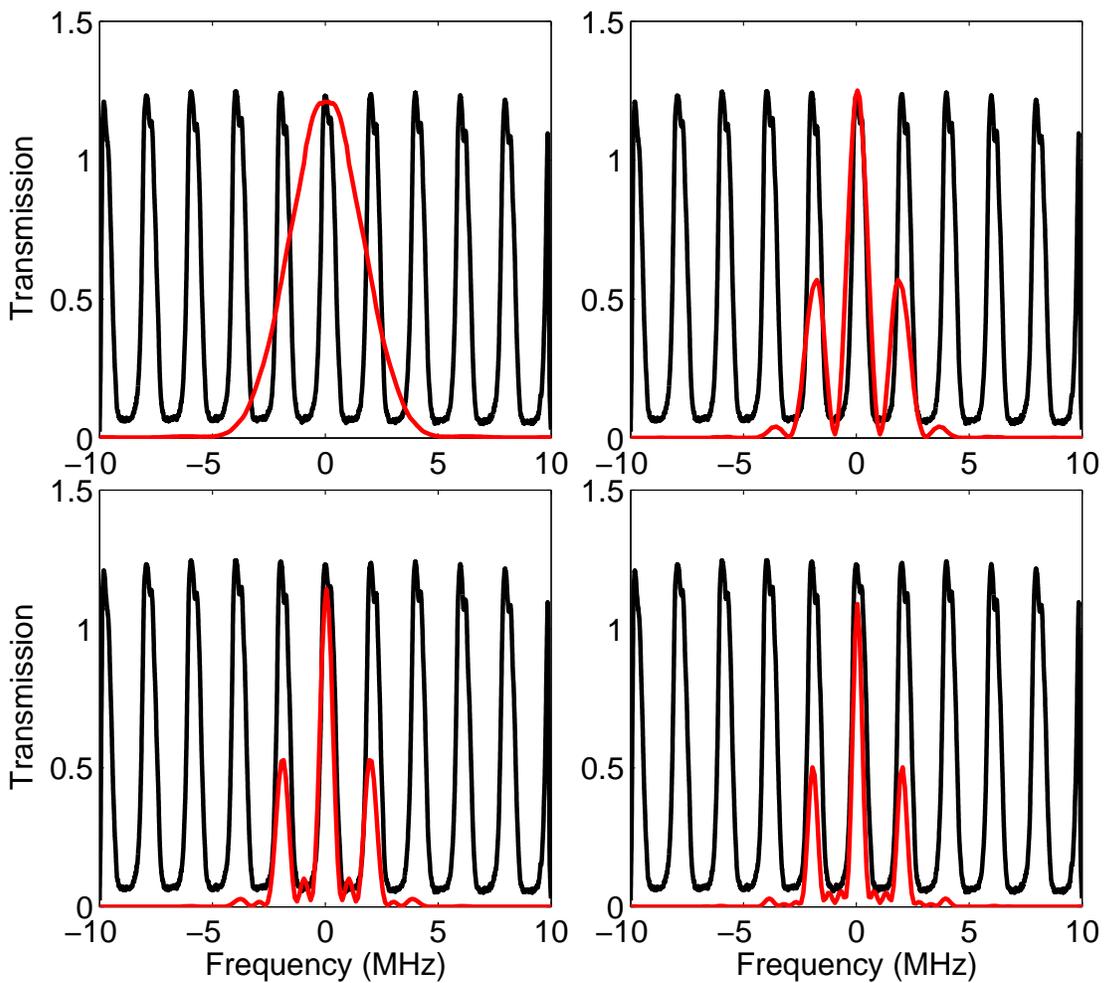}
	\caption{Comparison of the transmission spectrum in black (same as in. fig.\ref{peig}.a) with energy spectral density for one, two, three and four incoming pulses (a, b, c and d respectively in red). }	\label{figspectres}
	\end{center}
\end{figure}

With the same absorption spectrum as in fig.\ref{peig}.a, we now increase the number of incoming probe pulses. The pulses are precisely separated by $T=500$ns. As a consequence, the incoming optical spectrum becomes $2\pi /T$-periodic fitting the dispersive transparency windows of the comb. To illustrate this, we compare in fig. \ref{figspectres} the absorption spectrum (fig.\ref{peig}.a) obtained by our novel preparation method and the  energy spectral density for one, two, three and four incoming pulses ( fig. \ref{figspectres}.a, b, c and d respectively).

We indeed see that for an increasing number of pulses, the incoming spectrum is mostly restricted to the transparency windows of the comb. Dispersive effects should be apparent.

\begin{figure}[ht]
	\begin{center}
	\includegraphics[width=15cm]{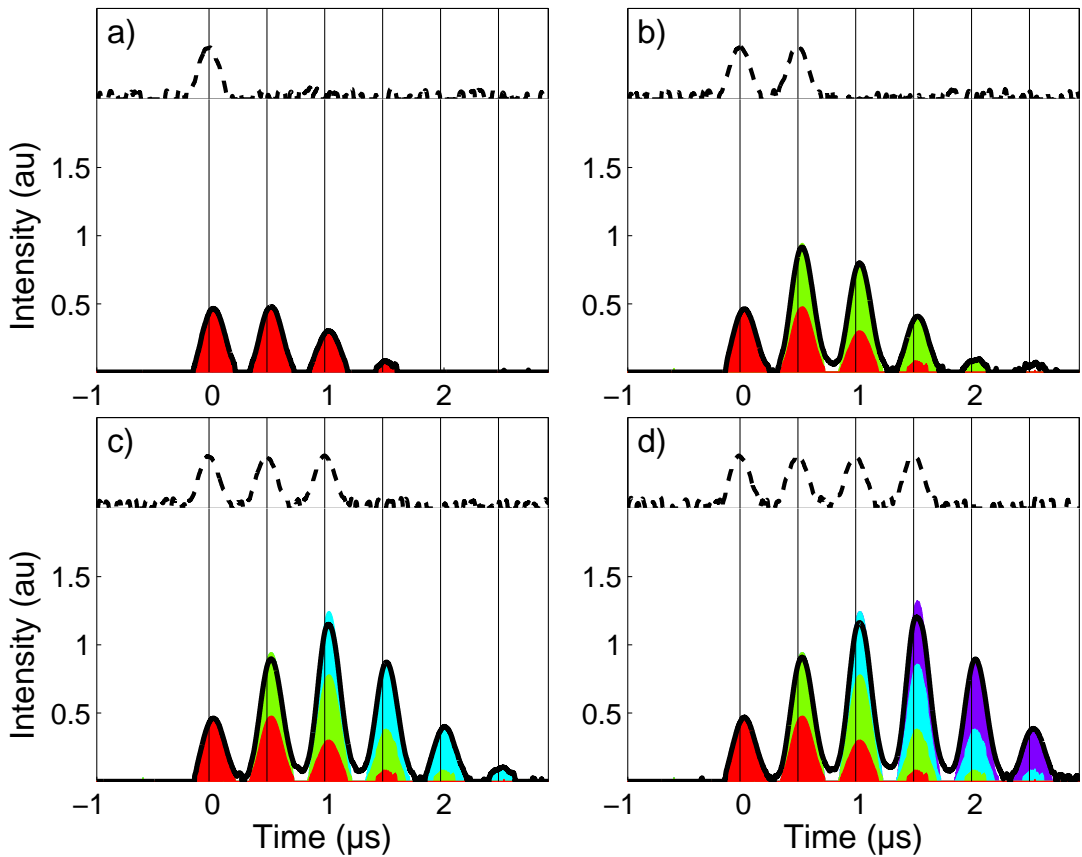}
	\caption{Echos of few incoming pulses. a) a single pulse is diffracted giving rise to a prime and secondary echos (similar to fig.\ref{peig}.b). The top dashed line represents the incoming pulses. b), c) and d) retrieval of two, three and four pulses respectively, separated by  $T=500$ns corresponding to the inverse comb period. See the text for details.}	\label{figTrain1234}
	\end{center}
\end{figure}

We can now investigate the exact same situation in the time domain with one, two, three and four incoming pulses  (fig.\ref{figTrain1234}.a, b, c and d respectively). It is particularly interesting because the observation can be alternatively interpreted as a series of echo diffracted by the different pulses or by the enhanced contribution of the dispersion.

Fig.\ref{figTrain1234} can be first interpreted as a superposition of successive echos from the incoming train. This is the goal of the colour code used to fit data. The echo of a single incoming pulse is taken as a reference (red area in fig.\ref{figTrain1234}.a). Then with two incoming pulses, we plot a red area for the echos of the first pulse and sum up the echos of the second pulse in green (fig.\ref{figTrain1234}.b). Even if we represent the intensity, the summation is done for the fields (square-root of the intensity). We apply the same procedure with three incoming pulses (the echo of the third pulse are in cyan in (fig.\ref{figTrain1234}.c) and four pulse (the echos of the fourth pulse are in magenta in (fig.\ref{figTrain1234}.d). The coloured areas are independently calculated from the single pulse echos. After summation, they should be compared to the black solid lines in fig.\ref{figTrain1234}.b), c) and d) which are the measured echos of two, three and four incoming pulses respectively. There is a good matching between the two showing that each outgoing pulse is actually composed of many echos generated by anterior incoming pulses. As an example in fig.\ref{figTrain1234}.d), the echo arriving at $t=1 \mu s$ is mostly composed of the transmitted part of the third pulse (0-order echo), plus the first echo of the second pulse, plus the second echo of the first pulse.

The outgoing pulses have a higher intensity than a single transmitted pulse due to the build-up of different echos. This statement can be easily interpreted in the spectral domain. By increasing the number of pulses, we actually restrict the incoming optical spectrum to the transparency windows of the comb. Avoiding the absorption peaks, the global transmitted intensity is larger and dispersion effect should be revealed. Even if the envelope of the outgoing train is not completely defined, we observe a delay by comparison to the incoming pulses (global shift of the envelope). This dispersive effect deserves further consideration. So we decide to send a much longer train with a well-defined smooth envelope. This regime should definitely reveal the dispersive nature of AFC.

\subsection{Echo of pulse train: slow-light regime}

\begin{figure}[ht]
	\begin{center}
	\includegraphics[width=8cm]{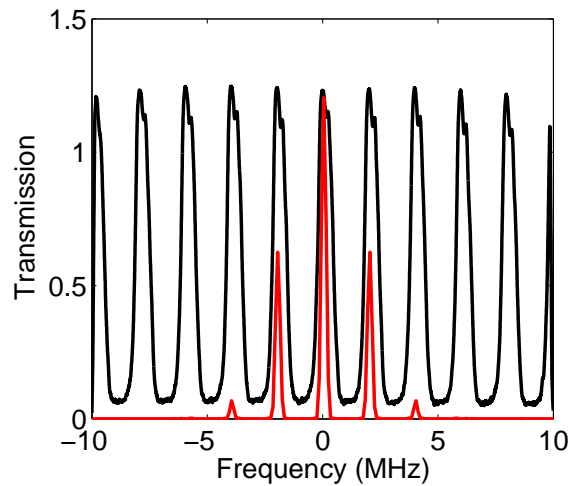}
	\caption{Energy spectral density (in red) of a gaussian train of pulses compared to atomic comb (in black). This incoming train will be use later on to measure the group delay (see fig. \ref{figTraingauss}). }	\label{figspectre10impEnvGauss}
	\end{center}
\end{figure}

We here use a gaussian shape for the train envelope. The resulting optical spectrum is then $2\pi /T$-periodic, gaussians in the spectral domain are well-restricted to the transparency regions of the comb as we can see in fig. \ref{figspectre10impEnvGauss}.

The incoming spectrum is now tightly restricted to the transparency region. As for any sort of transparent window, the dispersion should be linear to the first order to satisfy the Kramers-Kronig relation. As a consequence, we expect a global delay of the train envelope. Even if the effect was qualitatively observed in fig. \ref{figTrain1234}, an accurate measurement of the group delay requires a long train with well-defined envelope. This is the goal of the present section.

\subsubsection{Calculation of the AFC group delay from the absorption spectrum}\label{caldelay}

We use the same approach as in section \ref{sectionEcho}  by assuming a fully periodic absorption spectrum decomposed in a Fourier series $\displaystyle \alpha\left(\Delta\right)= \sum_n \alpha_n e^{i n \Delta T}$.

The dispersive effect can be derived from the Bloch-Maxwell equations written in the spectral domain
\begin{equation}
\begin{array}{ll}
\partial_z\widetilde{\Omega}\left(z,\omega\right)+ i \frac{\omega}{c} \, \widetilde{\Omega}\left(z,\omega\right) &=
-\displaystyle\frac{i}{2\pi} \widetilde{\Omega}\left(z,\omega\right) \int_\Delta \frac{\alpha \left(\Delta\right)}{\omega+\Delta-i\gamma} \mathrm{d} \Delta \\ [0.4cm]
&=
-\displaystyle\frac{i}{2\pi} \widetilde{\Omega}\left(z,\omega\right)  \sum_n \alpha_n \int_\Delta \frac{e^{i n \Delta T}}{\omega+\Delta-i\gamma} \mathrm{d} \Delta
\end{array}
\end{equation}

The Fourier transform of a complex lorentzian is calculated by using the Heaviside step function $Y$ (with $Y\left(0\right)=1/2$ \cite{bateman1954tables}). 

\begin{equation}
\begin{array}{ll}
\partial_z\widetilde{\Omega}\left(z,\omega\right)+ i \frac{\omega}{c} \, \widetilde{\Omega}\left(z,\omega\right) &=
-\displaystyle\ \widetilde{\Omega} \left(z,\omega\right)  \sum_n \alpha_n  e^{-i n \omega T} Y\left(n\right)  e^{- n \gamma T} \\[0.4cm]
&=-\displaystyle\ \frac{1}{2} \widetilde{\Omega} \left(z,\omega\right)  \left(\alpha_0+ 2\sum_{n>0} \alpha_n  e^{-i n \omega T} e^{- n \gamma T} \right)
\end{array}
\end{equation}
The group velocity $V_g$ is defined by the term $i \omega$ deduced from a Taylor expansion of
$e^{-i n \omega T} =1{-i n \omega T}+...$
\begin{equation}\label{V_g}
\frac{1}{V_g}=\frac{1}{c}- T \sum_{n>0} n \Re\left( \alpha_n \right) e^{- n\gamma T}
\end{equation}
It should be noted that within a transparency window, the $ \alpha_n$ coefficients are dominantly negative rendering the group velocity positive and possibly much smaller than $c$.

By analogy with section \ref{sectionEcho}, we can now compare the expected group delay $T_g=L/V_g$ that we deduce from the absorption spectrum $\displaystyle \alpha\left(\Delta\right)$ and the envelope delay that is actually observed.

The sum over the Fourier coefficients is infinite, only truncated by the factor $e^{- n\gamma T}$ due to the homogeneous broadening. We could consider the no-damping limit $\gamma \rightarrow 0$ as soon as the $\displaystyle \alpha\left(\Delta\right)$ resolution is much broader than $\gamma$. The sum would not be truncated but still convergent. In practice, the convergence may not be satisfied numerically because of a limited experimental resolution generating noisy coefficient at large frequencies (large $n$). We then keep the factor $e^{- n\gamma T}$ minimizing the influence of unresolved high frequency components. In \TMYAG, we independently measure $2 \pi / \gamma = 35 \mu s$. This value will be used for numerical applications as a truncation.

\subsubsection{Group delay measurements}

\begin{figure}[ht]
	\begin{center}
	\includegraphics[width=15cm]{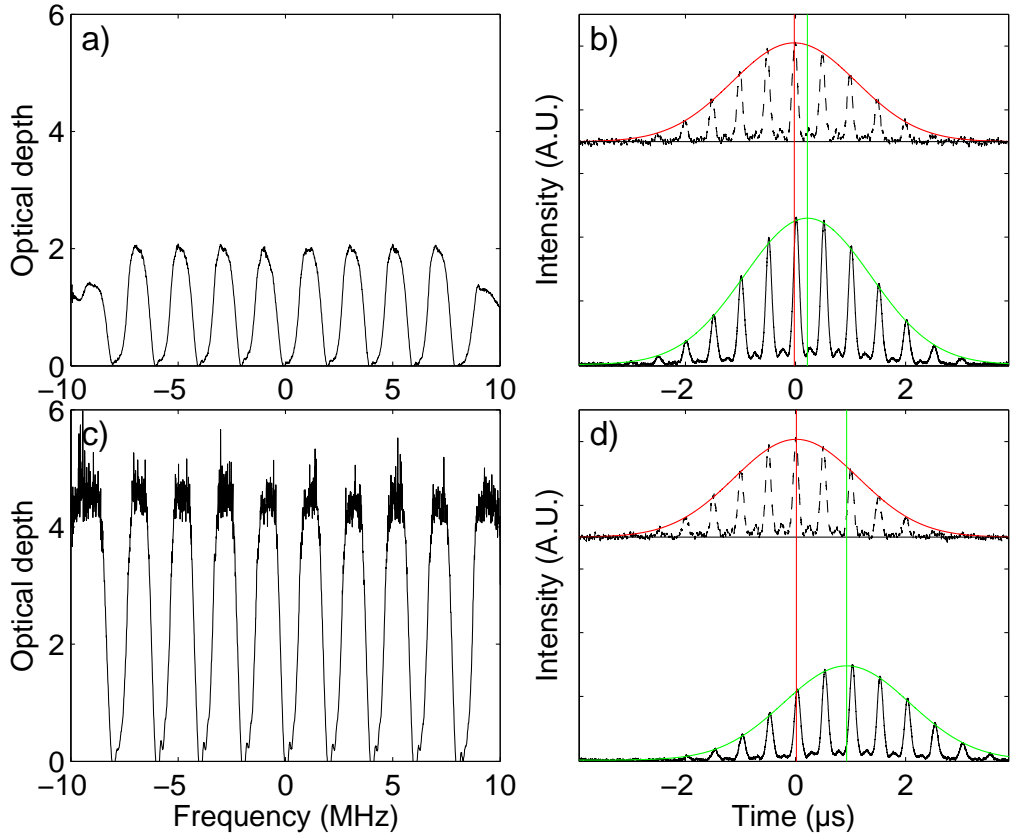}
	\caption{a) \& c) AFC obtained for different initial optical depths depending on laser central frequency within the inhomogeneous profile. b) \& d) delayed envelopes of the transmitted train (solid line under the green envelope) compared to the incoming pulse  (dashed line under the red envelope). The delays for b) \& d) are 240 ns \& 910 ns  {corresponding to 21km.s$^{-1}$ \& 5.5km.s$^{-1}$ for the group velocity} respectively (5mm crystal).}	\label{figTraingauss}
	\end{center}
\end{figure}

We here send a pulse train with a gaussian envelope through the AFC. We observe in fig.\ref{figTraingauss}.b. and d. (for different optical depths) that  the envelope is shifted by the group delay. To vary the maximum optical depth, we shift the laser central frequency within the inhomogeneous profile ($\sim$10GHz broad). We then obtain different combs (fig.\ref{figTraingauss}.a. and c) using the same preparation technique described before.

As expected, the higher the optical depth, the larger the delay.   {The largest observed delay is  910 ns corresponding to 5.5km.s$^{-1}$ for the group velocity. We typically retrieve the order of magnitude that we expect in our introductive analysis assuming a generic shape of the comb (see fig. \ref{figDisp}).}

As detailed in section \ref{caldelay}, the delay measurements can be directly compared to its calculation from the absorption spectrum. From the coefficients $\displaystyle \alpha_n $, we can calculate the group delay (eq. \ref{V_g}). We obtained from fig.\ref{figTraingauss}.a. \& c expected delays of 176 ns and 583 ns respectively.

\begin{figure}[ht]
	\begin{center}
	\includegraphics[width=10cm]{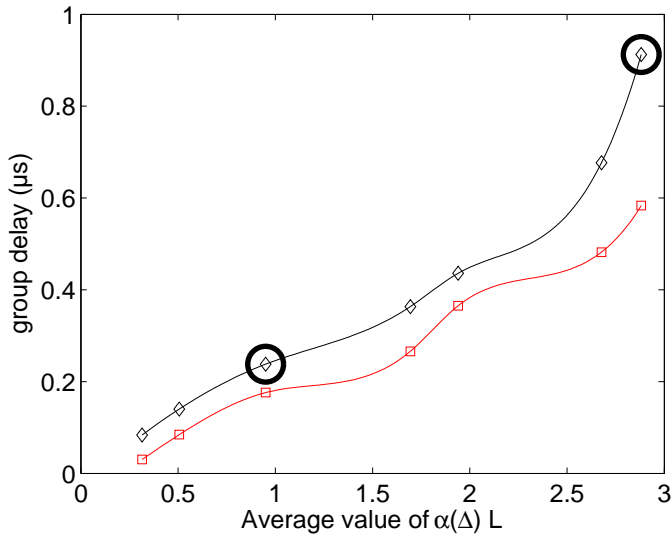}
	\caption{Measured delays of a gaussian pulse train for various initial optical depths (diamonds). It is compared with the group delay  (eq. \ref{V_g}) expected from the absorption spectrum (squares). We choose as abscissa the average optical depth that is easily accessible experimentally. Solid lines are used to guide the eye. The circled symbols correspond to the measurements of figs. \ref{figTraingauss}.b) \& d)}	\label{figcourberetard}
	\end{center}
\end{figure}

This comparison is repeated for different optical depths and is summarized in fig.\ref{figcourberetard}.
There is a global agreement  between the two values in fig.\ref{figcourberetard}. The difference is more significant at large optical depth. It emphasizes the difficulty to accurately measure weak transmission coefficients very sensitive to experimental artifacts \cite{AFC_NJP}. To conclude our analysis, we propose a classification of the protocols in terms of dispersion or absorption.

\section{Brief classification of the protocols}
The goal of the present work is to clarify the importance of dispersion in the AFC storage protocol. We do not intend to do an extensive review \cite[and references therein]{RevModPhys.82.1041, refId0} but we would like to place our analysis in a broader context. We propose to classify the different existing protocols in two categories related to absorption or dispersion depending on the dominant storage process. We hope this distinction will be the basis of future debates in the community.


\subsection{Dispersive storage protocols}
In this category, EIT is the historical example \cite{EITth}. The dispersive feature is given by a dip in the absorption profile \cite[and references therein]{lukinRMP} due to EIT in three-level $\Lambda$-system. An alternatively as been proposed based on spectral hole burning when a long-lived shelving state is present \cite{lauroslow}.

A steep dispersion dependency can also be obtained with off-resonance Raman excitation in a $\Lambda$-system \cite{NunnQM}. The so-called Raman memory can be treated by a universal formalism similar to EIT \cite{Gorshkov} but it should be noted that the dispersion curve is here significantly different \cite{refId0}.

We hope the reader is now convinced that the AFC \cite{AFCTh} belongs to this category as well.

Even if a dispersive effect can be observed in a wide range of situations, trapping light in the sample is a different story. Dispersion reveals the retarded response of the transition dipoles. The complete light storage is obtained by decoupling the dipoles from the field. This is essentially done by a Raman transfer  \cite{ lauroslow, Gorshkov} converting the transition dipoles into spin coherences. The retrieval corresponds to the inverse process.

\subsection{Absorbing storage protocols}
In this category they are all derived from the 2PE. The absorption is significant only if an inhomogeneous broadening is present because the process must occur faster than the coherence time. The optical spectrum is in that case larger than the homogeneous broadening but smaller than the inhomogeneous profile. The 2PE offers the possibility to retrieve light after absorption.

The 2PE is definitely an absorbing storage protocol but it is not well suited for quantum storage because of spontaneous emission \cite{ruggiero2PE, Sangouard2PE}. The CRIB solves this problem by applying static electric fields \cite{tittel-photon} and has been successfully applied in REDC \cite{Hedges2010}. More recently, an alternative has been proposed involving only optical excitations without spectral preparation \cite{ROSE, Hyper}.

For these protocols, the storage step is straightforward because it is based on direct absorption. The retrieval is more problematic because even if it is triggered by a rephasing of the coherences, the medium is still absorbing. Propagation effects can be detrimental at this stage. An appropriate control of the retrieval direction is required.

\section{Conclusion}
We have investigated the slow-light effect during the AFC storage. We essentially show that the propagation through the spectral comb is dominated by off-resonance excitation of the absorbing peaks. As a consequence the light storage is guided by dispersion. This statement is well-supported by group delay measurements in \TMYAG. We conclude by proposing a non-extensive classification of the different protocols. This analysis intends to give an overall view and should be placed in a broader context in order to gain a deeper understanding of the existing protocols and to conceive new ones.

\section*{Acknowledgments}

This work is supported by the European Commission through the FP7 QuRep project and by the national grant ANR-09-BLAN-0333-03. We thank the Universit{\'e} Paris-Sud-11 for supporting S.A. Moiseev during his visiting stay. We greatly appreciated the assistance of P. Goldner during experimental runs.

\section*{References}

\bibliographystyle{iopart-num}
\bibliography{AFCslow_JphysB_bib}

\end{document}